\documentclass[amsmath, amssymb, showpacs, reprint, floatfix, aps, prl, superscriptaddress]{revtex4-1}
\usepackage{graphicx}
\usepackage{dcolumn}
\usepackage{physics}
\usepackage{braket}
\usepackage{bm}
\usepackage{color}
\usepackage{hyperref}
\hypersetup{colorlinks=true,citecolor=blue,linkcolor=blue, urlcolor=blue}
\setcitestyle{super}
\newcommand{\kk}{\mathbf{k}}
\renewcommand{\qq}{\mathbf{q}}
\newcommand{\nk}{n\mathbf{k}}
\renewcommand{\vec}[1]{\mathbf{#1}}

\begin{document}
%
\author{Vatsal A. Jhalani$^{\dagger}$}
\affiliation{Department of Applied Physics and Materials Science, Steele Laboratory, California Institute of Technology, Pasadena, California 91125, USA.}
%
\author{Jin-Jian Zhou$^{\dagger}$}
\affiliation{Department of Applied Physics and Materials Science, Steele Laboratory, California Institute of Technology, Pasadena, California 91125, USA.}
%
\author{Marco Bernardi*}
\affiliation{Department of Applied Physics and Materials Science, Steele Laboratory, California Institute of Technology, Pasadena, California 91125, USA.}
\email{bmarco@caltech.edu}
%
\title{Ultrafast Hot Carrier Dynamics in GaN and its\\Impact on the Efficiency Droop}
%
%
\date{\today}
%
%
\begin{abstract}
* Corresponding author. Email: bmarco@caltech.edu\\
\vspace{10pt}
\hspace{1mm} $^\dagger$ These authors contributed equally to this work.\\
GaN is a key material for lighting technology. Yet, the carrier transport and ultrafast dynamics that are central in GaN light emitting devices are not completely understood. 
We present first-principles calculations of carrier dynamics in GaN, focusing on electron-phonon (e-ph) scattering and the cooling and nanoscale dynamics of hot carriers. 
We find that e-ph scattering is significantly faster for holes compared to electrons, and that for hot carriers with an initial 0.5$-$1 eV excess energy,  
holes take a significantly shorter time ($\sim$0.1 ps) to relax to the band edge compared to electrons, which take $\sim$1 ps. 
The asymmetry in the hot carrier dynamics is shown to originate from the valence band degeneracy, the heavier effective mass of holes compared to electrons, and the details of the coupling to different phonon modes in the valence and conduction bands. 
We show that the slow cooling of hot electrons and their long ballistic mean free paths (over 3 nm) are a possible cause of efficiency droop in GaN light emitting diodes.
Taken together, our work sheds light on the ultrafast dynamics of hot carriers in GaN and the nanoscale origin of efficiency droop.
\end{abstract}
\maketitle
%
%

\indent
Wurtzite GaN has emerged as a promising material for solid state lighting \cite{LEDProspects2009} and power electronics \cite{GaNPowerSwitching2015,PowerElectronics2015}, with potential technological benefits 
that are driving intense research in industry and academia. However, material properties essential for device performance and energy efficiency, such as carrier transport and recombination, are not completely understood in GaN and remain the subject of debate. 
%
Carrier transport and ultrafast dynamics are regulated by scattering with phonons, carriers and impurities \cite{Ziman}. 
In particular, the electron-phonon (e-ph) interaction \cite{Mahan-nutshell,Bernardi-review} plays a dominant role on transport at room temperature in relatively pure materials. 
It further regulates the energy loss (or \lq\lq cooling\rq\rq) of excited carriers injected at heterojunctions, a scenario of relevance in GaN light emitting diodes. 
The excited (so-called \lq\lq hot'') carriers rapidly lose their excess energy with respect to the band edges, dissipating heat by phonon emission through e-ph coupling.  
%
%
Hot carriers (HCs) are also central in degradation and current leakage in GaN transistors for power electronics \cite{MeneghessoHEMTReview2008,ChowdhuryVerticalReview2013}, and set the operational basis for hot electron transistors \cite{DasguptaHET2011}.\\
\indent
%
%
Microscopic understanding of carrier dynamics is challenging in GaN since experimental results are modulated by defects and interfaces, and are typically interpreted with empirical models \cite{Ye1999, Ye2000,Stanton2001,Balkan2002,Wang2006,Wu2007,Tripathy2008}.  
For example, the hot electron cooling times measured by different groups range over two orders of magnitude
\cite{Ye1999,Stanton2001,Balkan2002,Wang2006,Wu2007,Tripathy2008,Suntrup2014}, and reports of hot hole dynamics are scarce \cite{Ye2000}. 
In addition, the efficiency decline in GaN light emitting diodes (LEDs) at high current, a process known as efficiency droop \cite{VerzellesiDroopReview2013}, has been intensely investigated but its carrier dynamics origin remains unclear. 
First-principles calculations focused on Auger recombination \cite{Kioupakis2011,Kioupakis2015} as a possible cause, though other mechanisms have been proposed \cite{VerzellesiDroopReview2013}, including HC effects and electron leakage. 
These processes have seen less extensive theoretical treatment compared to Auger, leaving simplified models to guide \mbox{device design}.\\
\indent
%
%
We recently developed first-principles calculations of carrier dynamics \cite{Bernardi-review} that can obtain carrier mobility \cite{Mustafa,JinJianGaAs}, ultrafast dynamics \cite{MarcoSunlight,MarcoGaAs,Marco-SPP,Palummo}, HC relaxation times \cite{MarcoSunlight,MarcoGaAs} and ballistic mean free paths \cite{MarcoSunlight,Marco-SPP} in excellent agreement with experiment. These approaches are free of empirical parameters and use the structure of the material as the only input. 
In particular, we recently developed a method  \cite{JinJianGaAs} to accurately compute the e-ph relaxation times (RTs), namely the average time between e-ph collisions, in polar materials, as is needed for GaN. These approaches are extended in this work to investigate HC dynamics in GaN from first principles.\\
\indent
%
%
%
Here, we compute the e-ph RTs over a wide energy range, and study the cooling of HCs by numerically solving the electron Boltzmann transport equation (BTE).  
Both the RTs and the simulated ultrafast dynamics reveal a large asymmetry between the hot electron and hole dynamics, with hot holes relaxing to the band edges significantly faster than hot electrons. 
The origin of this asymmetry, the role of different phonon modes and the limitations and failure of phenomenological models are analyzed in detail. 
We additionally find significantly longer mean free paths (MFPs) for electrons compared to holes, with implications for GaN devices.  
We show that the slow cooling rate of hot electrons can lead to inefficient light emission at high current, thus demonstrating that the nanoscale dynamics of HCs play a key role in LED efficiency droop. \\
%
%
%
\indent \textbf{Electron-phonon scattering.} 
In polar materials like GaN, empirical models typically assume that polar optical phonons $-$ and in particular, the longitudinal optical (LO) mode in GaN $-$ dominate carrier scattering due to their long-range interactions with carriers. 
The empirical Fr{\"o}hlich model \cite{Frohlich} for the LO mode predicts an e-ph coupling matrix element \mbox{ $g_F (q) = [(e^2 \hbar \omega_{0} )/ (2 q)] \epsilon^{-1}_{\text{ph}}$ }, where $\qq$ is the phonon wavevector (and $q$ its magnitude), $\hbar \omega_{0}$ the LO phonon energy, and $\epsilon^{-1}_{\text{ph}} = \epsilon^{-1}_\infty - \epsilon^{-1}_0$ the phonon contribution to the dielectric screening, with $\epsilon_\infty$ and $\epsilon_0$ the high- and low-frequency dielectric constants, respectively. 
The intra-valley e-ph scattering rate $\Gamma_{\kk}$ (where $\kk$ is the electron crystal momentum) due to the empirical Fr{\"o}hlich coupling $g_F$ can be obtained analytically for carriers in a spherical parabolic band with effective mass $m^*$ and energy $E_{\kk} = \hbar^2 k^2 / 2m^*$ \cite{Ridley,Mahan-nutshell}:
%
%
\begin{equation} \label{eq:frohlich1}
\begin{aligned}
\Gamma_{\vec{k}}=&\,\tau_0^{-1}\!\left(\frac{E_{\kk}}{\hbar\omega_{0}}\right)^{\!\!-\frac{1}{2}}  \!\left[N_{0} \sinh^{-1}\!\left(\frac{E_{\kk}}{\hbar\omega_{0}}\right)^{\!\!\frac{1}{2}}\!  +  \right. \\
&\left. \left(N_{0}+1\right)\sinh^{-1}\!\left(\frac{E_{\kk}}{\hbar\omega_{0}}-1\right)^{\!\!\frac{1}{2}} \right]
\end{aligned}
\end{equation} 
where $N_0$ is the Bose-Einstein occupation factor for LO phonons and $\tau_{0}^{-1} = \epsilon^{-1}_{p} [(e^2 \omega_0 )/(2\pi \hbar )] \!\cdot\! [m^* / (2\hbar \omega_0)]^{1/2}$ is the inverse Fr{\"o}hlich time. 
%
In this work, the rate in \mbox{Eq.~\ref{eq:frohlich1}} is used to compare the widely employed Fr{\"o}hlich  empirical model with our first-principles results.\\
\indent
%
%
Following an approach we recently developed \cite{JinJianGaAs}, we combine density functional theory (DFT) \cite{Martin}, density functional perturbation theory (DFPT) \cite{Baroni} and \textit{ab initio} Fr{\"o}hlich \cite{Verdi-frohlich} calculations to compute the short- and long-range contributions to the e-ph coupling matrix elements, which are then interpolated on fine Brillouin zone (BZ) grids to converge the e-ph scattering rates $\Gamma_{\nk}$ for each electronic band $n$ and crystal momentum $\kk$ (see Methods). 
%
%
For the polar LO mode scattering, our \textit{ab initio} Fr{\"o}hlich calculations \cite{Verdi-frohlich} differ in important ways from the empirical Fr{\"o}hlich model as they include Born effective charges and anisotropic dielectric tensors, both computed with DFPT, and account for the electronic bandstructure and phonon dispersions (see Methods). Here and in the following, the carrier excess energies are defined as the energy above the conduction band minimum (CBM) for the electrons, and the energy below the valence band maximum (VBM) for the holes.\\
\indent
%
%
\begin{figure*}[!ht]
\includegraphics[scale=0.25]{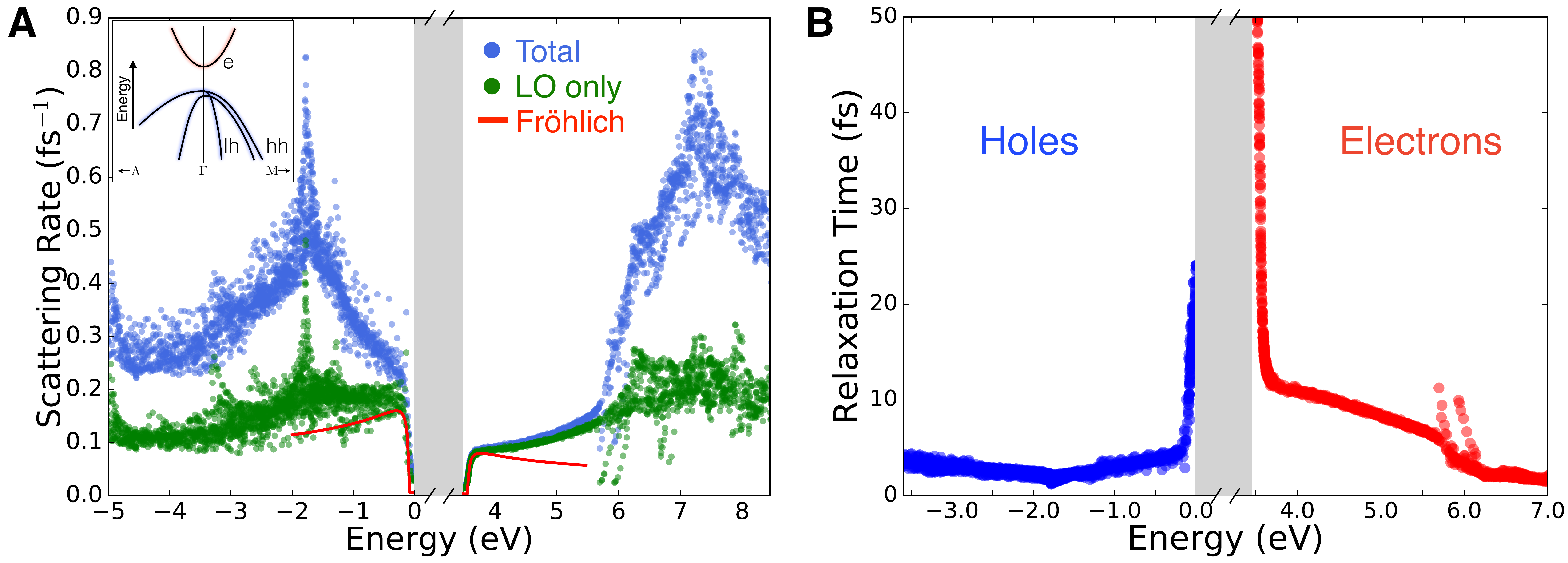}
\caption{\textbf{First-principles e-ph scattering.} (\textbf{A}) e-ph scattering rates in GaN at 300 K, for electrons and holes with energies within 5 eV of the band edges. Both the LO contribution and the total rate due to all phonon modes are shown. 
The red lines are the empirical Fr{\"o}hlich scattering rates (see Eq. \ref{eq:frohlich1}). A schematic of the bandstructure of GaN near the band edges is shown in the inset. (\textbf{B}) The e-ph relaxation times, defined as the inverse of the total scattering rates in (\textbf{A}). 
The zero of the energy axis is the valence band maximum, and the band gap is shown as a shaded area.
\label{fig1}}
\end{figure*}
%
%
The scattering rates and their inverse, the e-ph RTs $\tau_{\nk} = \Gamma_{\nk}^{-1}$, contain rich microscopic information on the carrier dynamics.
%
%
%
The bandstructure crucially determines the e-ph scattering rates. 
A schematic of the bandstructure of GaN near the band edges is given for reference in the inset of Figure \ref{fig1}A.
While the conduction band exhibits a single parabolic valley at $\Gamma$, the valence band edge consists of a light-hole and two heavy-hole bands with anisotropic effective masses and degeneracy along the $\Gamma$$-$A direction. 
Figure \ref{fig1}A shows our computed e-ph scattering rates of electrons and holes with energies within 5 eV of the band edges.  
Both the total scattering rate contributed by all phonons and the LO mode contribution alone are shown; the empirical Fr{\"o}hlich rate of \mbox{Eq.~\ref{eq:frohlich1}}, computed using parameters from the literature \cite{Pankove1975Properties, Bougrov2001Properties}, is also given for comparison. The full mode-resolved scattering rates are shown in Figure S1 of the Supplementary Materials. 
%
For both electrons and holes, the scattering rate in Figure \ref{fig1}A is very small within an LO phonon energy ($\hbar \omega_{0} \!\approx\! 100$ meV) of the band edges, since the dominant process in this energy range is LO phonon absorption. 
The scattering rate increases rapidly up to $\sim$150 meV excess energy due to an increase in the phase space for LO phonon emission above $\hbar \omega_{0}$. 
The trend at low energy in the conduction band is consistent with the conventional picture, with LO mode emission dominating intra-valley scattering in the conduction valley at $\Gamma$, roughly up to 2 eV above the CBM. 
At higher energy in the conduction band, the total and LO mode electron scattering rates differ substantially due to inter-band scattering mediated by all phonon modes in roughly equal measure (see Figure S1).\\
%
\indent
%
In the valence band, due to the presence of multiple bands at $\Gamma$, both intra-valley and inter-valley small-$\qq$ transitions are possible within $\sim$1 eV of the VBM, 
thus resulting in a higher LO scattering rate (by roughly a factor of 2) than in the conduction band within 1 eV of the CBM. 
%
%
The LO contribution becomes roughly constant at hole excess energies greater than $\sim$0.2 eV. Different from the conduction band where LO phonon emission is the only active process, the total scattering rate keeps increasing for holes with excess energy above 0.2 eV due to intra-valley and inter-valley scattering contributed, in roughly equal measure, by all acoustic and optical phonon modes (see Figure S1). 
Within 2 eV of the band edges, the DOS in the valence band is greater than in the conduction band due to the presence of multiple valence bands and to the higher effective masses of holes compared to electrons. 
These effects result in scattering from phonon modes other than the LO becoming important at lower excess energies in the valence band due to a greater phase space for large-$\qq$ scattering. Overall, the stronger LO polar and non-polar contributions in the valence band result in significantly higher scattering rates for holes compared to electrons within 2 eV of the band edges. 
This asymmetry has important consequences for carrier dynamics.\\ 
\indent
%
%
The empirical Fr{\"o}hlich model shows two major limitations in reproducing the first-principles trends (see Figure \ref{fig1}A). 
The empirical and first-principles rates exhibit opposite trends at excess energies greater than $\sim$150 meV.  
For energies where the LO phonon emission rate is roughly constant, the discrepancy of the empirical Fr{\"o}hlich rate is as large as 30$-$50\% for both electrons and holes. 
Note that first-principles calculations include all phonon modes on the same footing at all energies, while empirical e-ph calculations would account for modes other than the LO through mode-specific empirical deformation potentials \cite{Ridley}. 
Traditionally employed empirical models are thus inadequate to compute the e-ph scattering rates due to all phonon modes over a wide energy range, as is done here, and the first-principles approach is necessary.\\
\indent
%
As a consequence of the scattering rate asymmetry, the computed e-ph RTs (see Figure \ref{fig1}B) of holes are overall significantly shorter than the RTs of electrons. 
Within 2 eV of the band edges, the electron RTs range between 10$-$50 fs, while the hole RTs are of order 3$-$20 fs. 
The electron RT above the threshold for LO phonon emission in the conduction band is $\sim$12 fs, in very good agreement with the LO phonon emission time of 16 fs recently measured for electrons by Suntrup et al. \cite{Suntrup2014}. 
The detailed energy dependence of the RTs and scattering rates reported here is valuable for GaN device design.\\ 
%
\indent \textbf{Origin of the carrier relaxation asymmetry.} We address the question of whether the asymmetry between the electron and hole scattering rates found here is a mere consequence of the heavier effective mass of holes compared to electrons in GaN. In doing so, we develop an intuition for the origin of this asymmetry by analyzing separately the polar and non-polar e-ph scattering contributions.  
As noted above, the two sources of e-ph scattering are the long-range interaction from the LO polar mode and the short-range interactions from all other non-polar phonons.  
%
%
Because e-ph processes are determined by the e-ph coupling strength and the phase space available for scattering (see Methods, Eq.~\ref{eq:tau}), the non-polar scattering rate $\Gamma^{\text{(NP)}}$ approximately follows the same energy trend as the electronic DOS, $D(E)$, multiplied by an average e-ph coupling strength $\langle g^2 \rangle$, so that $\Gamma^{\text{(NP)}}(E) \!\propto\! \langle g^2 \rangle D(E)$. 
%
On the other hand, due to the long-range electrostatic nature of the polar interaction, the strength of LO coupling is insensitive to the specific electronic states involved in the scattering process.
The LO mode polar coupling behaves as $|g^{LO}(q)|^{2} \!\sim\! 1/q^{2}$ at small phonon wavevector $q$, resulting in much stronger LO scattering for small-$\qq$ transitions. 
Due to this particular phonon wavevector dependence, the scattering rate $\Gamma^{\text{(P)}}$ due to the polar LO mode approaches the band edge with a constant trend in energy, as opposed to being proportional to the DOS as in the non-polar case.    
In the empirical Fr{\"o}hlich formula (see Eq.~\ref{eq:frohlich1}), the polar scattering rate behaves as a simple function of the effective mass, $\Gamma^{\text{(P)}} \propto (m^*)^{1/2}$.\\
\indent
%
%
%
\begin{figure}[!t]
\includegraphics[scale=0.45]{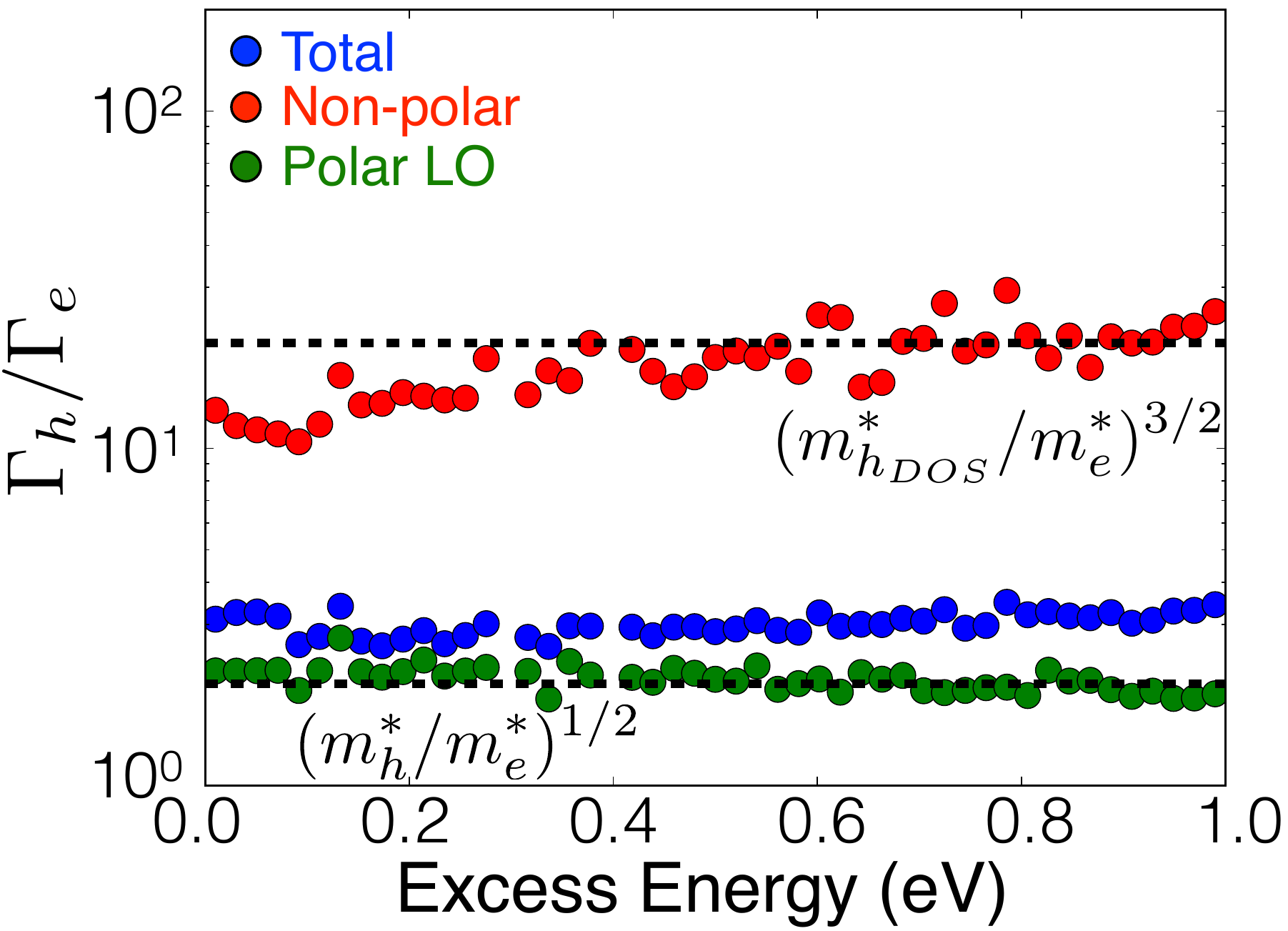}
\caption{\textbf{Origin of the scattering rate asymmetry.} 
\mbox{Ratio} of the Brillouin zone averaged e-ph scattering rates of holes ($\Gamma_h$) to those of electrons ($\Gamma_e$) as a function of carrier \mbox{excess energy.} 
The zeros of the excess energy are the conduction and valence band edges for electrons and holes, respectively. The data points are computed using rates due to polar LO phonons (green), non-polar phonons (red), and all phonon modes (blue). 
The dashed lines indicate the ratios $(m^{*}_{h}/m^{*}_{e})^{1/2}$ and $(m^{*}_{h_{DOS}}/m^{*}_{e})^{3/2}$ discussed in the text. 
\label{fig2}}
\end{figure}
To quantify the scattering rate asymmetry, we compute in Figure \ref{fig2} the ratio of the BZ averaged scattering rate of holes ($\Gamma_h$) to that of electrons ($\Gamma_e$). 
The ratio $\Gamma_h / \Gamma_e$ is shown for the average total, polar and non-polar scattering rates as a function of excess energy. 
For the LO polar contribution, the ratio closely matches the result expected based on the empirical Fr{\"o}hlich model (see Eq.~\ref{eq:frohlich1}), namely, $\Gamma^{\text{(P)}}_h/\Gamma^{\text{(P)}}_e\!\approx\!(m^{*}_{h}/m^{*}_{e})^{1/2}$, where $m^{*}_{h}$ is the experimental hole effective mass \cite{Pankove1975Properties}. The agreement between the empirical and \textit{ab initio} ratios of LO mode scattering rates indicates that the heavier hole effective mass is the main source of the LO polar scattering asymmetry, and that the empirical and \textit{ab initio} treatments roughly factor out in the polar scattering rate ratio when a single $m^*_h$ value is employed as a proxy of the multiple hole bands.\\
\indent
Since the non-polar scattering rate is proportional to the DOS, the ratio between the \textit{ab initio} non-polar scattering rates is compared in Figure \ref{fig2} with a heuristic DOS ratio. 
Since the DOS of a parabolic band is proportional to $(m^*)^{3/2}$ \cite{AshcroftMermin}, we expect that the ratio of the non-polar scattering rates at low energy is approximately   
$\Gamma^{\text{(NP)}}_h/\Gamma^{\text{(NP)}}_e\!\approx\! (m^{*}_{h_{DOS}}/m^{*}_{e})^{3/2}$, 
where $m^{*}_{h_{DOS}}$ is the hole DOS effective mass \cite{Bougrov2001Properties}. 
The \textit{ab initio} and DOS-based empirical non-polar ratios are in reasonable agreement. However, the inaccuracy of approximating multiple valence bands with a single DOS effective mass for holes, combined with the stronger average e-ph coupling strength for electrons (see Figure S2 and Section A in the Supplementary Materials), both push the ratio below the heuristic prediction, with an energy-dependent discrepancy.
Therefore, the ratio between the non-polar scattering rates cannot be accurately estimated without detailed knowledge of the bandstructure and e-ph coupling strengths.\\
\indent
Finally, since the total scattering rate is determined by the sum of the polar and non-polar contributions, the ratio between the \textit{total} scattering rates of holes and electrons cannot be estimated by a simple heuristic model based on the effective masses. 
The average \textit{ab initio} ratio between the total scattering rates found here is $\Gamma_h / \Gamma_e \!\approx\! 3$ within 1 eV of the band edges, and is bracketed by the non-polar and polar ratios. 
Detailed first-principles calculations, as employed here, are necessary to quantify this asymmetry in the scattering rates.\\
%
%
\begin{figure*}[!ht]
\includegraphics[scale=0.35]{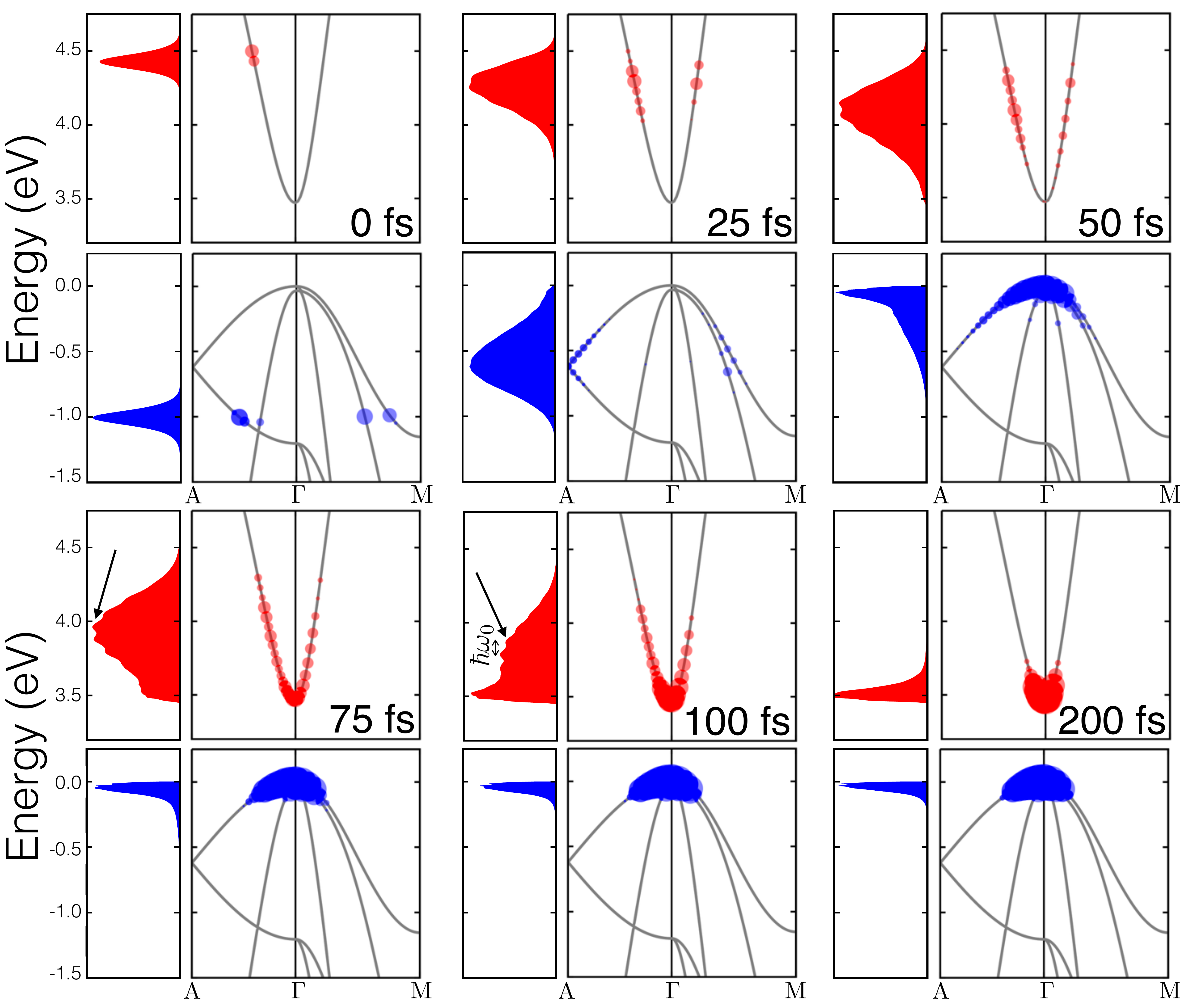} 
\caption{\textbf{Ultrafast dynamics of hot carriers in GaN.} Hot electrons (blue) and holes (red) are injected with a 1 eV excess energy with respect to the band edges, as modeled by creating initial Gaussian carrier distributions with a small energy width. The occupations $f_{n\kk}$ for electrons and $1 - f_{n\kk}$ for holes are shown at various times in the square panels along the A$-$$\Gamma$$-$M line of the Brillouin zone, with the point size proportional to the logarithm of the occupations. Left of each occupation panel we plot the average carrier concentrations (in arbitrary units) as a function of energy. The fluctuations in the electron concentration are indicated with an arrow in the 75 fs and 100 fs frames, along with their spacing of $\hbar \omega_0 \approx 100$ meV.
\label{dynamics}}
\end{figure*}
%
%
\indent \textbf{Real-time hot carrier dynamics.} 
Simulating the HC dynamics in real time further highlights the different behavior of electrons and holes in GaN. 
To this end, we carry out numerical simulations of the dynamics of hot electron and hole populations injected in GaN with a range of initial excess energies. 
To represent the injected carriers, we employ narrow Gaussians distributions centered at the initial excess energy, and solve the electron BTE \cite{Mahan-nutshell} in real time fully \textit{ab initio} \cite{Bernardi-review}, using first-principles e-ph matrix elements, bandstructures and phonon dispersions (see Methods). The carrier occupations are time-stepped using a 4$^{\text{th}}$-order Runge-Kutta algorithm, while the phonon occupations are kept at 300 K, so that hot phonon effects are neglected.\\
\indent
%
%
Figure \ref{dynamics} shows the time evolution of the electron and hole populations after injection with a 1 eV initial excess energy. At each time step, we analyze both the carrier occupations ($f_{n\kk}$ for electrons and $1 \!-\! f_{n\kk}$ for holes) along the A$-$$\Gamma$$-$M line of the BZ and the carrier concentrations as a function of energy, $\overline{f}(E)$, obtained by integrating the occupations at each energy over the BZ (see Methods). 
Following HC generation, the electron and hole distributions broaden in energy and approach a Fermi-Dirac-like shape as they shift toward lower excess energies, eventually reaching the band edges. 
Both electrons and holes ultimately thermalize to a 300 K Fermi-Dirac distribution in equilibrium with the phonons, thus reaching the correct long-time limit for our simulations.\\
\indent
We find that while holes reach the band edges and cool in $\sim$80 fs, electrons are still far from equilibrium at the same time.  
Electron cooling is roughly five times slower, with electrons relaxing to the band edges in $\sim$200 fs after injection, and fully thermalizing to the 300 K equilibrium distribution in 400 fs. 
Interestingly, the dominant LO phonon emission for electrons results in distribution fluctuations at energies spaced apart by $\hbar\omega_{0}\!\approx\!100$ meV. 
These wiggles in the distributions are seen most clearly for electrons in the 75 fs and 100 fs panels in Figure \ref{dynamics}. The same simulation carried out for an initial carrier excess energy of 0.5 eV (see Figure S3 in Supplementary Materials) shows even more pronounced fluctuations in the electron population. 
The asymmetry found here in the time scale for hot electron and hole cooling has not been reported previously, and is distinct from the transport asymmetry \cite{VerzellesiDroopReview2013} due to the different effective masses and mobilities of electrons and holes.\\ 
\indent
Comparing our real-time dynamics with experiments is not straightforward, and should be done keeping in mind that our results pertain to the low carrier concentration regime. 
Several ultrafast pump-probe experiments have been carried out in GaN, but due to the high HC concentrations reached in some of these measurements, 
hot phonon effects become relevant as the emitted phonons re-excite the carriers, thus slowing down HC cooling compared to the low carrier concentration regime studied here. 
%
For example, Ye et al. \cite{Ye1999} measured hot electron cooling in $n$-doped GaN. They concluded that LO emission is the dominant mechanism for hot electron cooling, consistent with our findings.
%
They extracted a 0.2 ps LO emission time from their measurements, and observed that this value is significantly longer than the $\sim$10 fs empirical Fr{\"o}hlich RT; 
this discrepancy is attributed in their work to hot phonon effects resulting from the high carrier density. 
%
In a subsequent work, the same authors measured the cooling of hot holes in $p$-doped GaN \cite{Ye2000}. They found that the hot hole data is difficult to fit with an LO phonon emission model because of the complexity of the valence band. We argue that the reason for this discrepancy, as shown in our work, is the significant scattering by phonon modes other than the LO in the valence band, even at low excess energy. 
While differences in the electron and hole dynamics are not discussed by Ye et al. \cite{Ye2000}, note that in their work the details of the early time evolution cannot be seen due to the 100 fs duration of the pump pulse, which sets the time resolution of the measurement. 
%
Our data provides evidence for an asymmetry between the electron and hole cooling times and mechanisms, which should be observable in new experiments with $\sim$10 fs resolution. \\
%
%
%
%
\begin{figure*}[!ht]
\includegraphics[scale=0.5]{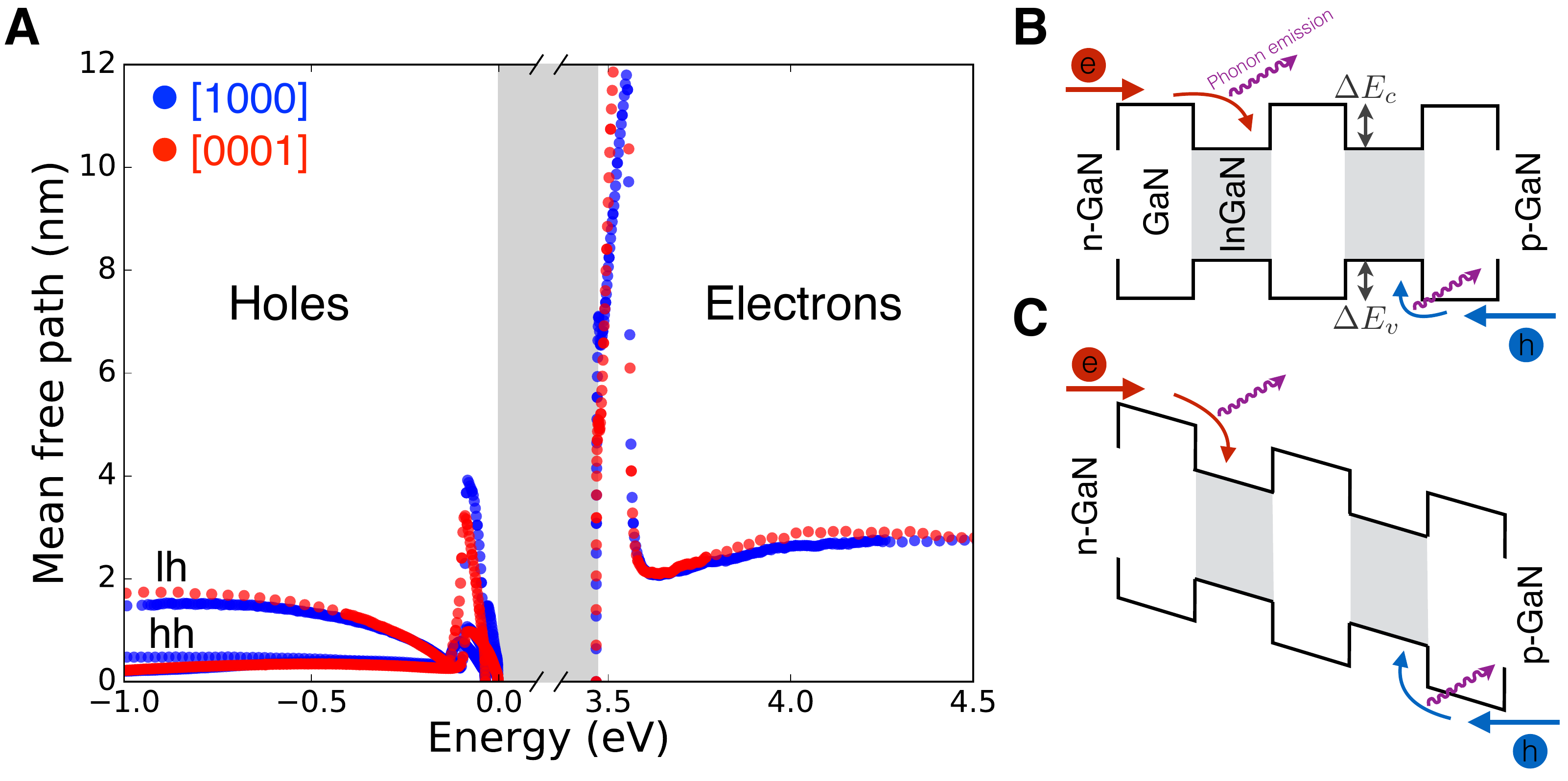} 
\caption{\textbf{Role of hot carriers in efficiency droop.} (\textbf{A}) Mean free paths, computed using calculated band velocities and e-ph relaxation times, along two crystal directions in GaN. 
The c-axis direction is labeled as [0001] and the in-plane direction as [1000]. For holes, the lower mean free path branch is due to the heavy-hole (hh) band, and the upper branch to the light-hole (lh) bands. 
The zero of the energy axis is the valence band maximum, and the band gap is shown with a shaded area. 
(\textbf{B})-(\textbf{C}) Schematic band diagram of the active region of a InGaN/GaN LED. Shown are the stacked quantum wells, 
both without and with a driving electric field, in (\textbf{B}) and (\textbf{C}) respectively. The valence and conduction band offsets, $\Delta E_{v}$ and $\Delta E_{c}$, are also indicated.
\label{fig4}}
\end{figure*} 
%
%
\indent \textbf{Role of hot carriers in efficiency droop.}
We analyze the role of HCs in GaN LED devices. 
GaN and III-nitride LEDs are bipolar devices that emit light as a result of the recombination of electrons and holes in quantum well (QW) active layers \cite{LEDProspects2009,Hardy-review}. 
A key figure of merit in LEDs is the external quantum efficiency (EQE), defined as the ratio of generated photons per electron-hole pair introduced at the contacts. 
%
%
One of the main barriers toward commercialization of GaN LEDs is efficiency droop, a decay of the EQE at high driving current. 
Almost all GaN LEDs exhibit a maximum EQE at low current densities, with the EQE rapidly decreasing above the optimal current \cite{VerzellesiDroopReview2013}.\\ 
%
\indent
In a typical GaN LED design \cite{LEDProspects2009}, electrons and holes are injected at the opposite ends of an active region that contains multiple InGaN QWs and has the bandstructure shown schematically in Figure \ref{fig4}B. 
The multi-QW active region typically consists of up to 10 periods of a $\sim$3 nm thick In$_x$Ga$_{1-x}$N QW followed by an undoped $\sim$10 nm GaN spacer \cite{VerzellesiDroopReview2013}, a structure designed to maximize carrier confinement and recombination in the light emitting InGaN QWs. 
Due to the band offsets at the GaN/InGaN interface, electrons and holes are hot when injected into the QW, and need to cool to the band edges to emit light efficiently. 
%
The excess energy of the HCs at injection in the QW depends on the band offsets $\Delta E_c$ for electrons and $\Delta E_v$ for holes (see Figure \ref{fig4}B), which are typically of order 0.5 eV for InGaN/GaN devices.\\
\indent
Carriers injected into the active region scatter with phonons, and after several phonon emission events lose enough energy to be confined in the QW by the band offset potential, 
following which the carriers cool to the QW band edges and recombine to emit light. 
The e-ph mean free paths (MFPs) $L_{n\kk}$ characterize the average length traveled between phonon emission events by carriers moving in the direction $\hat{\vec{k}}$ in band $n$ (here, the top valence bands for holes and bottom conduction band for electrons). 
The MFPs are obtained here as $L_{n\kk} = v_{n\kk}\tau_{n\kk}$, using Wannier-interpolated band velocities $v_{n\kk}$ \cite{WannInterp1, WannInterp2} and the e-ph relaxation times $\tau_{n\kk}$ given in Figure \ref{fig1}B.  
Figure \ref{fig4}A shows the computed e-ph MFPs for carriers moving along the c-axis [0001] and in-plane [1000] directions in GaN, for energies within 1 eV of the band edges, which are typical for carrier injection in GaN LEDs. 
%
The slow hot electron cooling via LO emission results in relatively long electron MFPs ranging from 10 nm near the CBM to 3 nm between 0.25$-$1 eV above the CBM.  
For comparison, the hole MFPs are much shorter, $\sim$3 nm close to the VBM and only 0.5 nm for the heavy-hole and 1.5 nm for the light-hole bands between 0.25$-$1 eV below the VBM. 
Note that these MFPs are lower bounds that pertain to the low carrier concentration regime, and that hot phonon effects can increase these values at high carrier concentration, especially for electrons where the only active e-ph scattering mechanism is LO phonon emission.\\
\indent
%
Since the typical QW thickness is $\sim$3 nm, the longer electron MFPs indicate that electrons need to cross on average several QWs before cooling to the band edge, while holes can cool effectively within a single QW. 
A fact relevant for droop is that as the current is increased, 
%
%
the driving electric field increases and further tilts the band edges in the active region, resulting in the tilted staircase potential shown in Figure \ref{fig4}C. 
For typical LEDs operated at droop regime currents, using the devices in Ref.~\cite{Droop-Tanaka} as an example, the driving voltage is roughly 5 V. 
Assuming that $\sim$3 V fall over a roughly 150 nm thick undoped active region, the resulting active region electric field is of order $2\times10^5\text{ V/cm}$.  
For electrons losing an LO phonon energy of $\sim$100 meV as they travel a 3 nm MFP length, the cooling rate is of order 100 meV over 3 nm, and thus roughly $\!3\times10^5 \text{ eV/cm}$. 
Droop thus occurs in a regime where the band tilting and the electron cooling rate are comparable, so that electrons cannot cool effectively against the tilted band edges. 
In this scenario, electrons that do not cool in the QW are swept by the electric field into the next QW, until a fraction of electrons leaks out of the active region without recombining radiatively, 
leading to a decreased EQE at high current as observed in efficiency droop. Our analysis highlights that HC cooling, 
and in particular the slow LO phonon emission rate of hot electrons occurring over lengths comparable with the QW thickness, can be an important factor determining efficiency droop.\\
%
%
\indent
The mitigation of droop seen experimentally by increasing the QW thickness \cite{Droop-Gardner} and number of QWs in the active region \cite{Droop-Tanaka} is consistent with our findings. 
In our model of hot electron leakage, increasing the QW thickness or the total active region thickness will improve HC cooling and decrease the probability of electrons overflying the QWs, thus increasing the EQE. 
Interestingly, the above reports attribute the droop mitigation primarily to reduced Auger recombination rather than HC effects.  
While Auger processes may play a role in droop, experimental evidence for Auger recombination in GaN LEDs is still scarce \cite{Zhang-review,VerzellesiDroopReview2013}. 
Auger processes, if present with a high enough rate, would need to coexist with the e-ph scattering processes studied here, which are relevant at all carrier concentrations. 
Most notably, it has been found that the mechanism responsible for droop increases in strength at lower temperatures \cite{Hader2011}. This behavior is in contrast with Auger scattering, since the dominant phonon-assisted Auger processes \cite{Kioupakis2015} would lead to stronger droop at higher temperatures. Note instead that e-ph scattering increases with increasing temperatures, so that lower temperatures are associated with longer hot carrier MFPs. 
Our model thus predicts that lower temperatures lead to longer electron MFPs and thus increased electron leakage and efficiency droop, consistent with experiment.  
The temperature dependence clearly suggests that e-ph scattering plays a key role in the efficiency droop.\\
\indent
Several authors have discussed HC effects and QW overfly as a potential source of droop \cite{VerzellesiDroopReview2013,Ni2010HotElecLED}, showing that the use of electron blocking layers to stop carrier leakage and staircase injectors to lower carrier excess energy can drastically mitigate droop. 
Our first-principles calculations of e-ph scattering, HC cooling and MFPs provide a quantitative basis for understanding such HC effects.  
We conclude that the different cooling rates of hot electrons and holes found here, together with the longer electron MFPs compared to holes, can play an important role in GaN LEDs efficiency droop and deserve further experimental investigation. \\
%
%
%
%
\indent
In summary, we apply first-principles calculations to shed light on the microscopic HC dynamics in GaN, providing details beyond the reach of previously employed theoretical and experimental methods. 
Our key findings include a significant difference in hot hole and electron scattering and cooling, the origin of this asymmetry, and the non-negligible role of scattering mechanisms besides the polar LO mode for holes.  
A model is presented to explain droop as a consequence of hot electron cooling. Our results can be employed to include HC dynamics in GaN device scale modeling, which typically neglects HC effects. 
Future work will extend our calculations to InGaN and AlGaN. 
Taken together, our computational approach advances the design of GaN devices, and enables the engineering of novel lighting materials with microscopic insight.\\
%
%
%
%
%
\indent
{\bf Methods.} {\it Computational methods.} We carry out \textit{ab initio} calculations on GaN in the wurtzite structure with relaxed lattice parameters of $a\!=\! 3.17$\,\AA~and $c\!=\!5.16$\,\AA.  
The ground-state electronic structure is computed using the {\sc Quantum Espresso} code~\cite{QE} within the local density approximation (LDA)~\cite{PZ} of DFT.
We employ a plane-wave kinetic energy cutoff of 80~Ry and scalar-relativistic norm-conserving pseudopotentials~\cite{TM} for both Ga and N. 
The pseudopotential of Ga includes a non-linear core correction~\cite{NLCC} to account for the effect of the shallow 3\textit{d} core states. The ground-state charge density is obtained using a $12\!\times\!12 \!\times\! 12$ $\kk$-point grid, 
following which a non-self-consistent calculation is employed to obtain the Kohn-Sham eigenvalues and wavefunctions on an $8 \!\times\! 8 \!\times\! 8$ $\kk$-point grid. We construct maximally localized Wannier functions (WFs)~\cite{Wannier-1} using the Wannier90 code~\cite{Wannier-2}. 
The Kohn-Sham wavefunctions are first projected onto four $sp^3$ orbitals on each Ga and N atom, for a total of 16 wannierized bands.  
The WF spread is then minimized, and the relevant energy windows~\cite{Wannier-1} are adjusted until the interpolated bandstructure can smoothly reproduce the LDA result within $\sim$10 meV throughout the BZ. 
The LDA eigenvalues are then corrected using the GW \cite{Hybertsen} quasiparticle energies of GaN computed in Ref. \cite{GaNRubioGW}, 
by extracting an average energy dependent self-energy that is employed to apply a scissor shift of the conduction band minimum and a 
linear energy stretch of the valence and conduction band energies. Using the GW-corrected eigenvalues together with the WFs, interpolated GW quasiparticle energies with quality comparable to those in Ref. \cite{GaNRubioGW} are obtained, and employed in all calculations described below. 
The WF projected DOS in Figure S2 of the Supplementary Materials is computed by projecting the Kohn-Sham wavefunctions $\ket{\psi_{n\kk}}$ onto the WFs localized on the Ga and N atoms, respectively, using $\sum_{\nk, \alpha \in \textrm{\{Ga, N\}}}\innerproduct{\psi_{nk}}{\alpha}\innerproduct{\alpha}{\psi_{nk}}\delta(E-E_{\nk})$.\\
%
%
\indent {\it Electron-phonon scattering calculations.}
We use density functional perturbation theory (DFPT)~\cite{Baroni} to compute lattice dynamical properties and e-ph matrix elements $g_{nn'\nu}(\kk,\qq)$ \cite{Bernardi-review} on coarse $8\!\times\!8 \!\times\! 8$ $\kk$-point and $4 \!\times\! 4 \!\times\! 4$ $\qq$-point grids in the BZ. 
Here and in the following, the e-ph matrix elements \cite{Bernardi-review} describe an electron in Bloch state $\ket{n\mathbf{k}}$, with quasiparticle energy $E_{n\mathbf{k}}$, that scatters into the state $\ket{n'\mathbf{k}+\mathbf{q}}$ with quasiparticle energy $E_{n'\mathbf{k}+\mathbf{q}}$, due to a phonon with branch index $\nu$, wavevector $\mathbf{q}$ and frequency $\omega_{\nu \mathbf{q}}$. 
The electron and phonon energies and the e-ph matrix elements are then interpolated on significantly finer grids using WFs~\cite{EPW-PRB}.
Wannier interpolation of the matrix elements relies on the spatial localization of the e-ph interaction.  
In polar materials like GaN, however, the interaction of an electron with an LO phonon diverges as $\sim$$1/q$ as predicted by the Fr{\"o}hlich model~\cite{Frohlich, Verdi-frohlich}. This singularity for $q \!\rightarrow\! 0$ results from the long-range field generated by the LO mode and is incompatible with Wannier interpolation. 
To overcome this issue, the e-ph matrix elements are separated into short- ($S$) and long-range ($L$) parts ~\cite{Verdi-frohlich}, $g_{nn'\nu}(\kk,\qq) = g_{nn'\nu}^S(\kk,\qq)+g_{nn'\nu}^L(\kk,\qq)$, where $g^{S}$ is computed by Wannier interpolation and $g^{L}$ is evaluated using an analytical \textit{ab initio} Fr{\"o}hlich e-ph vertex based on the Vogl model~\cite{Vogl1976}:
\begin{equation}\label{eq:gL}
\begin{split}
g^{L}_{nn'\nu}(\kk,\qq) =&\, i \frac{4\pi}{\Omega} \frac{e^2}{4\pi\epsilon_0} 
\sum_{\kappa} \qty(\frac{\hbar}{2\omega_{\nu\qq} N M_\kappa})^\frac{1}{2} \\
& \times \sum_{\vec{G}\neq -\qq} 
\frac{\qty(\qq+\vec{G})\cdot Z_\kappa^* \cdot \vec{e}_{\kappa\nu}\qty(\qq)}{\qty(\qq+\vec{G})\cdot \boldsymbol{\epsilon}^\infty \cdot \qty(\qq+\vec{G})} \\
& \times \mel{\Psi_{n'\kk+\qq}}{e^{i\qty(\qq+\vec{G}) \cdot \qty(\vec{r}-\boldsymbol{\tau}_\kappa)}}{\Psi_{n\kk}},
\end{split}
 \end{equation}
where $\vec{G}$ is a reciprocal lattice vector, $\Omega$ is the volume of the unit cell, $N$ is the number of points in the $\qq$-grid, $\boldsymbol{\epsilon}^\infty$ is the high-frequency permittivity, and $M_\kappa$, $\tau_\kappa$, and $Z_\kappa^*$ are the mass, position, and Born effective charge tensor of atom $\kappa$ in the unit cell, respectively, and $\vec{e}_{\kappa\nu}\qty(\qq)$ is a vibrational eigenmode normalized in the unit cell. 
Note that we neglected a quadrupole moment contribution to polar phonon scattering due to the piezoelectric interaction \cite{Vogl1976, JinJianGaAs}; we have verified that this approximation has a negligible effect on the hot carrier dynamics.\\
\indent
The e-ph scattering rate $\Gamma_{n\kk}$ for an electronic state with band $n$ and crystal momentum $\kk$ is obtained from the imaginary part of the lowest-order e-ph self-energy, using \cite{Bernardi-review,Mahan-nutshell}: 
\begin{equation} \label{eq:tau}
\begin{split} 
\Gamma_{n\kk}=&\frac{2\pi}{\hbar} \sum_{n' \qq \nu } \abs{g_{nn'\nu}(\kk,\qq)}^{2} \times \\
  & [\qty(N_{\nu\qq}+1-f_{n'\kk+\qq})\, \delta(E_{\nk} - \hbar\omega_{\nu\qq}-E_{n'\kk+\qq})\\
  &  + \qty(N_{\nu\qq}+f_{n'\kk+\qq})\, \delta(E_{\nk} +\hbar\omega_{\nu\qq}-E_{n'\kk+\qq})], 
\end{split}
\end{equation}
where $E_{n\kk}$ and $\hbar\omega_{\nu\qq}$ are the electron quasiparticle and phonon energies, respectively, and $f_{\nk}$ and $N_{\nu\qq}$ the corresponding equilibrium occupations at 300 K. The scattering rate due to a given phonon mode is obtained by restricting the sum in Eq.~\ref{eq:tau} to the corresponding phonon branch index $\nu$. 
%
Using an approach we recently developed~\cite{JinJianGaAs}, we compute and converge the scattering rate in Eq.~(\ref{eq:tau}) with an in-house developed code that can carry out, among other tasks, e-ph calculations analogous to the {\sc EPW} code~\cite{EPW}. 
We split $|g|^{2}$ in Eq.~(\ref{eq:tau}) into the long-range part $|g^{L}|^{2}$ and the remainder ($|g|^{2}-|g^{L}|^{2}$), and separately compute and converge $\Gamma_{n\kk}$ for these two contributions, which are then added up to obtain the total scattering rate.  This approach leads to a dramatic speed-up compared to converging Eq.~(\ref{eq:tau}) with $|g|^2\!=\!|g^S + g^L|^{2}$ directly \cite{JinJianGaAs}.
The scattering rate is computed for \mbox{$\kk$ points} on a fine $100 \!\times\! 100 \!\times\! 100$ grid in the BZ, using Gaussian broadening with a small parameter of 8 meV to approximate the $\delta$ function in Eq.~(\ref{eq:tau}).  
For each \mbox{$\kk$ point}, we converge the long-range contribution with $10^{6}$ random \mbox{$\qq$ points} sampled from a Cauchy distribution~\cite{JinJianGaAs}, 
and the remainder contribution with $10^{5}$ random \mbox{$\qq$ points} from a uniform distribution. Convergence of all quantities is carefully verified.\\
%
%
%
\indent {\it Carrier dynamics simulations.} We simulate HC cooling due to e-ph scattering in bulk GaN in the absence of external fields.  
The initial HC distribution is modeled with a narrow Gaussian centered at the initial HC energy. 
Separate calculations are carried out for electrons and holes, and for the two values of initial excess energy studied here (0.5 eV and 1 eV). 
The time evolution of the carrier distributions is obtained by solving the BTE~\cite{Bernardi-review, Mahan-nutshell}:  
\begin{equation} \label{eq:bte1}
\begin{aligned} 
\frac{\partial{ f_{n\kk}(t) }}{ \partial{t} }\!=\!& -\frac{2\pi}{\hbar} \!\sum_{n'\qq \nu} \abs{g_{nn'\nu}(\kk,\qq)}^{2}[ \delta(E_{\nk} \!-\! \hbar\omega_{\nu\qq} \!-\! E_{n'\kk+\qq}) \\
 &  \times F_{\text{em}}(t) +\delta(E_{\nk} + \hbar\omega_{\nu\qq} - E_{n'\kk+\qq})F_{\text{abs}}(t)],
 \end{aligned}
\end{equation}
where $f_{n\kk}(t)$ is the time-dependent electron distribution, the e-ph matrix elements $g_{nn'\nu}(\kk,\qq)$ are computed in the ground state, 
and the phonon emission ($F_{\text{em}}$) and absorption ($F_{\text{abs}}$) terms are computed at each time step as~\cite{Bernardi-review}: 
\begin{equation}\label{eq:bte2}
\begin{aligned}
F_{\text{abs}}& = f_{\nk}(1-f_{n'\kk+\qq})N_{\nu\qq} - f_{n'\kk+\qq}(1-f_{\nk})(N_{\nu\qq}+1) \\
F_{\text{em}}& = f_{\nk}(1-f_{n'\kk+\qq})(N_{\nu\qq}+1) - f_{n'\kk+\qq}(1-f_{\nk})N_{\nu\qq}.
\end{aligned}
\end{equation}
Phonon-phonon scattering and the change in phonon occupations are neglected in our simulations, where we fix $N_{\nu\qq}$ to the equilibrium phonon occupations at 300~K. 
This approximation is justified in the low carrier concentration limit, in which hot phonon effects are negligible. 
We solve Eq.~(\ref{eq:bte1}) numerically using the 4th-order Runge-Kutta method with a time step of 1 fs, using uniform fine grids of up to $100 \!\times\! 100 \!\times\! 100$ $\kk$- and $\qq$-points in the BZ. 
The BZ averaged energy-dependent carrier populations $\bar{f}(E,t)$ at energy $E$ used in Figure \ref{dynamics} are obtained as $\bar{f}(E, t) \!=\! \sum_{\nk}{ f_{\nk}(t)\delta(\epsilon_{\nk} - E)}$ via tetrahedron integration. 
We developed an efficient scheme to speed up the solution of the BTE, to be detailed elsewhere, 
which combines MPI and OpenMP parallelizations to repeatedly compute the e-ph matrix elements and converge the scattering integral at each time step.\\

%
%
\noindent
{\bf \large Acknowledgements}
\begin{acknowledgements}
V.J. thanks the Resnick Sustainibility Institute at Caltech for fellowship support. J.-J.Z. acknowledges support from the Joint Center for Artificial Photosynthesis, a DOE Energy Innovation Hub, as follows: the development of the computational methods employed in this work was supported through the Office of Science of the U.S. Department of Energy under Award No.~DE-SC0004993. M.B. acknowledges support by the National Science Foundation under Grant No. ACI-1642443, which provided for basic theory and part of the electron-phonon coupling code development. This research used resources of the National Energy Research Scientific Computing Center, a DOE Office of Science User Facility supported by the Office of Science of the U.S. Department of Energy under Contract No. DE-AC02-05CH11231. The authors thank Davide Sangalli for fruitful discussions.\\
\end{acknowledgements}

\noindent
{\bf \large Author Contributions}\\
M.B. conceived and designed the research. V.J. and  J.-J.Z. developed the computational codes and carried out the calculations. All authors wrote the manuscript. \\

\noindent
{\bf \large Supplementary Materials}\\
Supplementary material for this article is available at (URL to be added by the editorial office).
\noindent Section A. Origin of the stronger average e-ph coupling strength for electrons compared to holes.\\
\noindent Fig.~S1. Phonon mode-resolved electron-phonon scattering rates. \\
\noindent Fig.~S2. Origin of the stronger average electron-phonon coupling strength for electrons compared to holes.\\
\noindent Fig.~S3. Simulated hot carrier dynamics for 0.5 eV initial excess energy.\\

\noindent
{\bf Competing financial interests:} The authors declare no competing financial interests.\\

\noindent
\bibliography{ref}
\end{document}